\documentclass[aps,prd,twocolumn,showpacs,10pt,superscriptaddress,preprintnumbers,nofootinbib]{revtex4-2}
\usepackage{amsmath,amssymb,bm}\allowdisplaybreaks[1]
\usepackage{graphicx}
\usepackage{physics}
\usepackage{xcolor}
\usepackage[colorlinks, linkcolor=blue!80!black, citecolor=green!50!black, urlcolor=magenta]{hyperref}
\usepackage[normalem]{ulem}
\usepackage{adjustbox}
\usepackage{tikz}
\usetikzlibrary{arrows.meta}

\newcommand{\fig}[1]{Fig.~\ref{#1}}
\newcommand{\eq}[1]{Eq.~\eqref{#1}}
\newcommand{\eqs}[2]{Eqs.~\eqref{#1} and~\eqref{#2}}
\newcommand{\refcite}[1]{Ref.~\cite{#1}}
\newcommand{\refs}[1]{Refs.~\cite{#1}}

\newcommand{\pp}[1]{\left(#1\right)}
\newcommand{\bb}[1]{\left[#1\right]}
\newcommand{\cc}[1]{\left\{#1\right\}}

\newcommand{\beq}[1][]{\begin{equation}\label{#1}}
\newcommand{\eeq}{\end{equation}}
\newcommand{\bse}[1][]{\begin{subequations}\label{#1}}
\newcommand{\ese}{\end{subequations}}
\newcommand{\nn}{\nonumber}

\renewcommand{\P}{\mathcal{P}}
\newcommand{\wt}[1]{\widetilde{#1}}
\newcommand{\sgn}[1]{{\rm sgn}\left[#1\right]}

\newcommand{\M}{\mathcal{M}}

\newcommand{\N}{\mathcal{N}}
\newcommand{\F}{\mathcal{F}}

\newcommand{\Ct}{\widetilde{C}}

\newcommand{\GeV}{{\rm GeV}}

\begin{document}
%%%%%%%%%%%%%%%%%%%%%%%%%%%%%%%%%%%%%%%%%%%%%%%%%%%%%%%%

\preprint{
	{\vbox {			
		\hbox{\bf JLAB-THY-24-3985}
		\hbox{\bf MSUHEP-24-001}
}}}
\vspace*{0.2cm}

%================================================================================
\title{Extracting transition generalized parton distributions from hard exclusive pion-nucleon scattering}
%================================================================================

%--------
\author{Jian-Wei Qiu}
\email{jqiu@jlab.org}
\affiliation{Theory Center, Jefferson Lab,
Newport News, Virginia 23606, USA}
\affiliation{Department of Physics, William \& Mary,
Williamsburg, Virginia 23187, USA}
%--------

%--------
\author{Zhite Yu}
\email{yuzhite@jlab.org (corresponding author)}
\affiliation{Theory Center, Jefferson Lab,
Newport News, Virginia 23606, USA}
\affiliation{Department of Physics and Astronomy, 
Michigan State University, East Lansing, Michigan 48824, USA}
%--------

\date{\today}
%================================================================

%================================================================
\begin{abstract}
We study the extraction of transition generalized parton distributions (GPDs) from production of two back-to-back high transverse momentum photons ($\gamma\gamma$) and a massive pair of leptons ($\ell^+\ell^-$) in hard exclusive pion-nucleon scattering.  We argue that the exclusive scattering amplitude of both processes could be factorized into nonperturbative pion distribution amplitude and nucleon transition GPDs that are convoluted with perturbatively calculable short-distance matching coefficients.  We demonstrate that the exclusive diphoton production not only is complementary to the Drell-Yan-type dilepton production for extracting the GPDs, but also provides enhanced sensitivities for extracting the parton momentum fraction $x$ dependence of the GPDs.  We show that both exclusive observables are physically measurable at the J-PARC and AMBER experiment energies.  If the target nucleon can be polarized, corresponding spin asymmetries can offer additional sensitivities for extracting transition GPDs.  
\end{abstract}
%================================================================

\maketitle

%================================================================
\section{Introduction}
\label{sec:intro}
%================================================================

With a high energy charged $\pi$ beam at J-PARC~\cite{Aoki:2021cqa} and AMBER at CERN~\cite{Adams:2018pwt}, 
hard exclusive processes on a nucleon target ($N$) can provide 
unique opportunities to extract transition generalized parton distributions (GPDs), $\F_{NN'}(x, \xi, t)$, 
which could offer fundamental nonperturbative transition information from a hadron $N$ to $N'$.   
If $N$ and $N'$ are nucleons, the transition GPDs can be related to nucleon GPDs through 
isospin symmetry~\cite{Mankiewicz:1997aa, Frankfurt:1999fp}.  
Nucleon GPDs are nonperturbative hadron correlation functions which incorporate the concepts of 
parton distribution functions, distribution amplitudes (DAs), and form factors in their various aspects 
(for reviews, see~\cite{Goeke:2001tz, Diehl:2003ny, Belitsky:2005qn, Boffi:2007yc}). 
The Fourier transform with respect to $t$ at $\xi=0$ maps out the parton density distribution $f(x, \bm{b}_T)$ 
at a position $\bm{b}_T$ in the transverse plane perpendicular to the direction of the colliding hadron 
and in slices of different values of longitudinal parton momentum fraction $x$.  
The various $x$ moments, on the other hand, provide important information on the hadron's emergent properties
such as mass~\cite{Ji:1994av, Ji:1995sv, Lorce:2017xzd, Metz:2020vxd}, spin~\cite{Ji:1996ek},
and internal pressure and shear force distributions~\cite{Polyakov:2002yz, Polyakov:2018zvc, Burkert:2018bqq}.

Experimental extraction of GPDs of a hadron $h$ of momentum $p$ can be achieved by measurements of 
QCD factorizable single diffractive hard exclusive processes (SDHEPs)~\cite{Qiu:2022pla}
in its collisions with a beam particle $B(p_2)$,
$h(p) + B(p_2) \to h'(p') + C(q_1) + D(q_2)$,
where $h'(p')$ represents a diffractively produced hadron along with two (or more) exclusively produced particles $C(q_1)$ and $D(q_2)$ 
with a large transverse momentum $q_T \sim q_{1T} \sim q_{2T}$ that defines the hard scale of the collision.
Such processes can be thought of as occurring in two stages, 
\bse\label{eq:two-stage}\begin{align}
    h(p) \to &\, A^*(\Delta = p - p') + h'(p'), \\
    &\hspace{1ex} 
    \begin{tikzpicture}
        \node[inner sep=0pt] (arrow) at (0, 0) {
            \tikz{\draw[->, >={Stealth}, double, double distance=1pt, line width=1pt] (0, 0.28) to [out=-90, in=180] (0.8, 0);}
        };
    \end{tikzpicture}
    \hspace{1ex} \label{eq:second-stage}
    A^*(\Delta) + B(p_2) \to C(q_1) + D(q_2),
\end{align}\ese
where the diffractive subprocess is connected to the hard scattering through a virtual state 
$A^*$ of momentum $\Delta = p - p'$, which has a low invariant mass $t = \Delta^2$ 
but high energy, $\Delta^+ = 2 \xi P^+$, where $P = (p + p') / 2$,
and propagates for a long lifetime before entering the short-time hard interaction.
The two-stage picture provides a clear view of factorization, which necessarily requires $q_T \gg \sqrt{|t|}$.
Then the diffractive subprocess is factorized to 
a universal GPD function, separated from the hard interaction.
The latter, on the other hand, serves as a probe for the GPD to map out partonic information
of the transition from $h(p)$ to $h'(p')$.

Depending on the beam particle $B(p_2)$ and specific hard scattering subprocesses, the SDHEPs can probe various types of 
GPDs~\cite{Ji:1996nm, Radyushkin:1997ki, Brodsky:1994kf, Frankfurt:1995jw, Berger:2001xd, Guidal:2002kt, Belitsky:2002tf, Belitsky:2003fj, Kumano:2009he, Kumano:2009ky, ElBeiyad:2010pji, Pedrak:2017cpp, Pedrak:2020mfm, Siddikov:2022bku, Siddikov:2023qbd, Boussarie:2016qop, Duplancic:2018bum, Duplancic:2022ffo, Duplancic:2023kwe, Qiu:2023mrm}. 
With an electron beam, we have well studied electroproduction processes, such as
 deeply virtual Compton scattering (DVCS)~\cite{Ji:1996nm, Radyushkin:1997ki} 
and deeply virtual meson production (DVMP)~\cite{Brodsky:1994kf, Frankfurt:1995jw}.
With a photon beam, we have timelike Compton scattering (TCS)~\cite{Berger:2001xd}.
The pion (and also kaon) beams at J-PARC and AMBER at CERN allow 
``mesoproduction'' processes to probe transition GPDs,
which can be expressed in terms of flavor diagonal GPDs and are only sensitive to the valence parts of the latter.  
Two processes of this type have been identified:
the Drell-Yan (DY) dilepton production process~\cite{Berger:2001zn, Goloskokov:2015zsa, Sawada:2016mao},
\beq[eq:dilepton]
	N(p) + \pi(p_2) \to N'(p')+ \gamma^* [\to \ell^-(q_1) + \ell^+(q_2)],
\eeq
and the diphoton production process,
\beq[eq:diphoton]
	N(p) + \pi(p_2) \to N'(p') + \gamma(q_1) + \gamma(q_2),
\eeq
which was first introduced in \refcite{Qiu:2022bpq}.
See also \refs{Pire:2016gut, Pire:2021hbl, Pire:2022kwu} for discussions of pion beam use to probe transition distribution amplitudes.

For exclusive hard scattering like that in Eqs.~(\ref{eq:dilepton}) and (\ref{eq:diphoton}), QCD factorization requires at least two active partons from each identified hadron in the scattering amplitude due to the color singlet nature of the identified hadron. 
The hadron correlation functions associated with the factorization of exclusive processes are 
generally harder to extract from experimental data than their inclusive counterparts,
especially for their dependence on the relative momentum fraction of the two active partons, e.g., the $x$ dependence of GPDs.
This is because at the amplitude level, $x$ is fully integrated from $-1$ to $1$ in the factorization formulas.
Kinematically, owing to its exclusive nature, the invariant mass of the produced $C(q_1)$ and $D(q_2)$ 
of the $2\to 3$ SDHEPs in \eq{eq:two-stage} is uniquely determined by the observed $p'$.  
It is the distribution of the transverse momentum $q_T$ of $C(q_1)$ [or $D(q_2)$] in \eq{eq:second-stage} that could provide
additional details of the state $A^*(\Delta)$ and the $x$ dependence of the GPDs 
if the $x$ integration is entangled with the observed $q_T$ distribution.
However, such entanglement is absent from most processes, including the DVCS, DVMP, TCS, and DY dilepton production, 
where the $q_T$ dependence simply factors out of the $x$ dependence in the hard scattering, at least to the leading order, 
leaving only the GPD moment [see \eq{eq:GPD-moment}] to be measured.
Without considering the relatively weak QCD evolution effects~\cite{Moffat:2023svr},
extracting GPDs merely from such moments gives infinite families of solutions
quantified by the so-called shadow GPDs~\cite{Bertone:2021yyz}.

Fortunately, we could have processes~\cite{Qiu:2022bpq, Qiu:2023mrm} whose hard coefficients do not factorize 
in the above sense, allowing us to have more direct access to the hadron's partonic structure or the $x$ dependence of the GPDs.
As demonstrated in \refcite{Qiu:2023mrm}, such new processes can greatly 
reduce the ambiguity caused by shadow GPDs due to the entanglement 
between the measured hard scale and $x$ dependence of the GPDs. 

%----------------------------------------------------------------
% Fig: frame
%----------------------------------------------------------------
\begin{figure}[htbp]
	\centering
	\includegraphics[width=0.45\textwidth]{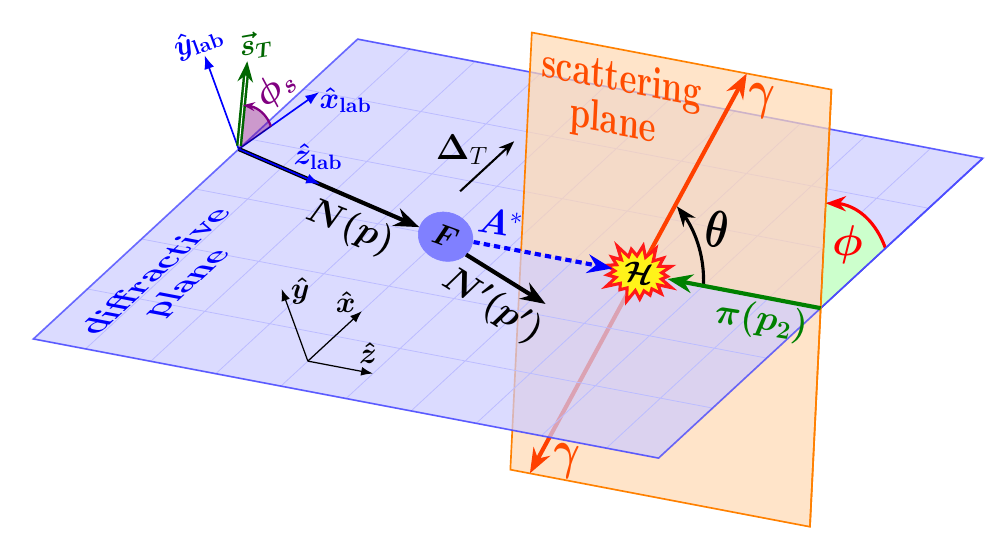}
	\caption{Diphoton mesoproduction process [\eq{eq:diphoton}] in the SDHEP frame. 
	The DY dilepton process [\eq{eq:dilepton}] is obtained by simply replacing the two final-state photons with leptons.}
	\label{fig:frame}
\end{figure}
%----------------------------------------------------------------

In this paper, we explore the role of SDHEPs with a charged $\pi$ beam for extracting transition GPDs  
while focusing on the diphoton production process in \eq{eq:diphoton}, where
the two observed photons do not come from the decay of a single virtual particle,
unlike the DY dilepton process in \eq{eq:dilepton}.  
We will evaluate both the production rate and the target spin asymmetry,
then explore how they can help distinguish the shadow GPDs from the real GPDs,
complementing the DY dilepton process at the same facilities.  
Our study could be extended further by replacing the two photons with two back-to-back high transverse momentum mesons, for example.

%================================================================
\section{Kinematics and observables}
\label{sec:kin}
%================================================================

%------------------------------------------------------------------------------------------------
\subsection{Kinematics and the two-stage paradigm}
\label{ssec:kin}
%------------------------------------------------------------------------------------------------

Following \eq{eq:two-stage} and \refs{Qiu:2022pla, Qiu:2023mrm}, 
we describe the kinematics of both mesoproduction processes in \eqs{eq:dilepton}{eq:diphoton}
in a two-stage manner, as shown in \fig{fig:frame}. 
First, the nucleon diffraction subprocess $N(p) \to N'(p') + A^*(\Delta)$ is described in the lab frame,
which is chosen as the center-of-mass (c.m.) frame of the colliding nucleon $N$ and pion $\pi(p_2)$, 
with the $\hat{z}_{\rm lab}$ being along the nucleon momentum $\bm{p}$ and 
$\hat{x}_{\rm lab}$ along the transverse momentum $\bm{\Delta}_T$ of the virtual state $A^*$.
This choice of $\hat{x}_{\rm lab}$ axis is not fixed in the lab frame but varies from event to event,
so does the direction of the transverse spin $\bm{s}_T$ of the nucleon $N$ with respect to $\hat{x}_{\rm lab}$, 
which has a varying azimuthal angle $\phi_S$ 
and effectively plays the role of the azimuthal angle of the diffractive plane.

Second, the hard scattering subprocess $A^*(\Delta) + \pi(p_2) \to \gamma(q_1) + \gamma(q_2)$ 
[or $\ell^-(q_1) + \ell^+(q_2)$] is described in the SDHEP frame, 
defined as their c.m.~frame, with $A^*$ moving along the $\hat{z}$ direction and 
$\hat{x}$ axis lying on the diffractive plane pointing in the same direction as the $\bm{\Delta}_T$ in the lab frame.
The kinematics of this hard scattering is quantified by its c.m.~energy squared,
$\hat{s} = (\Delta + p_2)^2 \simeq 2\xi s / (1 + \xi)$ [with $s = (p + p_2)^2$ the total collision energy squared],
the polar and azimuthal angles $(\theta, \phi)$ of one of the final-state particles.
For both processes, the cross section is symmetric with $\theta \to \pi - \theta$, 
so we may equivalently use the transverse momentum $q_T = (\sqrt{\hat{s}} / 2) \sin\theta$ in the place of $\theta$.
The SDHEP frame is in fact the same as the frames chosen in \refs{Berger:2001xd, Berger:2001zn},
except for the two-stage perspective adopted here.

%------------------------------------------------------------------------------------------------
\subsection{Factorization}
\label{ssec:fac}
%------------------------------------------------------------------------------------------------

As argued in \refs{Qiu:2022pla, Qiu:2022bpq}, the two-stage paradigm is materialized when $q_T \gg \sqrt{|t|}$
such that $A^*$ lives much longer than the timescale of its exclusive hard scattering with the pion.
With the quantum interference between these two stages suppressed by powers of $\sqrt{|t|} / q_T$, 
the dynamics taking place at these two stages could be factorized up to the entangled power corrections.
With only charged pion beams in experiments, the diffracted nucleon changes flavor, 
forbidding the simplest possibility of $A^*$ being a virtual photon. 
To the leading power of $\sqrt{|t|} / q_T$, $A^* = [q\bar{q}']$ is a collinear quark-antiquark pair,
and the diffractive subprocess is factorized from the whole amplitude into flavor transition GPDs.
Owing to the absence of a gluon channel ($A^* = [gg]$), the factorization breaking issue raised in \refcite{Nabeebaccus:2023rzr} does not apply here.

For the diphoton process, we have the factorization~\cite{Qiu:2022bpq}
\begin{widetext}
\begin{align}
	\mathcal{M}_{N \pi \to N' \gamma_{\lambda_1} \gamma_{\lambda_2}} 
	= \int_{-1}^{1} dx \int_0^1 dz \, D^{[q'\bar{q}]}_{\pi}(z) 
		\bb{ \wt{F}^{[q\bar{q}']}_{NN'}(x,\xi,t) \, C^{\gamma}_{\lambda_1\lambda_2} (x, \xi; z; \theta)
			+ F^{[q\bar{q}']}_{NN'}(x,\xi,t) \, \wt{C}^{\gamma}_{\lambda_1\lambda_2} (x, \xi; z; \theta)
		}, 
\label{eq:factorize}
\end{align}
\end{widetext}
where $\lambda_{1,2}$ are the photon helicities in the SDHEP frame.
$D^{[q'\bar{q}]}_{\pi}$ is the DA for the beam $\pi$ to turn into a collinear quark-antiquark pair $[q'\bar{q}]$.
$F^{[q\bar{q}']}_{NN'}$ and $\wt{F}^{[q\bar{q}']}_{NN'}$ are, respectively, unpolarized and polarized nucleon transition GPDs 
from $N$ to $N'$ for exchanging a $[q\bar{q}']$ pair with the hard collision, 
which can be decomposed into the GPDs $H$, $E$, $\wt{H}$, and $\wt{E}$,
and $\wt{C}^{\gamma}$ and $C^{\gamma}$ are the corresponding perturbative hard coefficients 
describing the exclusive scattering
of the two collinear and on-shell quark-antiquark pairs into the two back-to-back photons, 
\begin{equation}
	[q\bar{q}'](\hat{p}_1) + [\bar{q}q'](\hat{p}_2) \to \gamma(q_1) + \gamma(q_2),
\end{equation}
with $\hat{p}_1 = (\Delta \cdot n)\bar{n}$, $\hat{p}_2 = (p_2 \cdot \bar{n})n$, and lightlike vectors
$\bar{n}^\mu=\delta^{\mu +}$ and ${n}^\mu=\delta^{\mu -}$ in the SDHEP frame.

Owing to the pseudoscalar nature of pion and chiral symmetry at leading power, 
quark transversity GPDs do not contribute, so
the scattering in \eq{eq:factorize} has no net helicity in the initial state.
The resultant hard coefficients $C^{\gamma}$ and $\wt{C}^{\gamma}$ contain no nontrivial $\phi$ dependence.
The detailed calculations were presented in \refcite{Qiu:2022bpq} 
in terms of gauge-invariant tensor decomposition,
and are reproduced in Appendix \ref{app:diphoton} as helicity amplitudes of the photons.

For the DY dilepton process,
to the leading order of QED, the lepton pair arises from the decay of a timelike photon $\gamma^*$
that is produced via the annihilation of the two quark pairs in \eq{eq:dilepton}.
Ward identity and parity symmetry constrains the $\gamma^*$ to be longitudinal and 
indicates that only the polarized GPD contributes, 
\begin{align}
	\mathcal{M}_{N \pi \to N' \ell_{\lambda_1} \ell_{\lambda_2}} 
	& = \int_{-1}^{1} dx \int_0^1 dz \, D^{[q'\bar{q}]}_{\pi}(z) 
	\nn\\
	&\hspace{1.5em} \times
		\wt{F}^{[q\bar{q}']}_{NN'}(x,\xi,t) \, C^{\ell}_{\lambda_1\lambda_2} (x, \xi; z; \theta),
\label{eq:dilepton-factorize}
\end{align}
where for comparison with the diphoton production, the hard coefficient $C^{\ell}_{\lambda_1\lambda_2}$ is presented
in terms of the polar angular distribution of one of the observed leptons in Appendix \ref{app:dilepton}.

In this paper, we consider two reaction channels. 
For the $\pi^-$ beam, we have $(NN') = (pn)$ and $(q q') = (u d)$,
and for the $\pi^+$ beam, we have $(NN') = (np)$ and $(q q') = (d u)$.
In both cases, it is the transition GPD 
\begin{align}
	\mathcal{F}^{[u\bar{d}]}_{pn} = \mathcal{F}^{[d\bar{u}]}_{np} = \mathcal{F}^u_p - \mathcal{F}^d_p
\end{align}
(with $\mathcal{F} = F$, $\wt{F}$, $H$, $E$, $\wt{H}$, or $\wt{E}$)
that is probed.
In the following, unless explicitly indicated, the GPDs refer to such flavor specification.

%------------------------------------------------------------------------------------------------
\subsection{Cross section and single target spin asymmetry}
\label{ssec:obs}
%------------------------------------------------------------------------------------------------

The polarization observables are greatly restricted by the scalar nature of the pion beam.
Without the ability to measure polarizations of final-state particles, 
parity limits the only spin asymmetry to be $\sin\phi_S$
associated with the transverse nucleon spin $\bm{s}_T$~\cite{Klem:1976ui, Bunce:1976yb, Sivers:1989cc}, 
constructed out of $\bm{p} \cdot (\bm{\Delta}_T \times \bm{s}_T)$ in the lab frame for the diffractive subprocess.
Integrating out the trivial $\phi$ dependence in the hard scattering
results in the differential cross section
\begin{align}
	&\frac{d \sigma}{d |t| \, d\xi \, d\cos\theta \, d\phi_S} 	\nn\\
	&\hspace{1em}
		= \frac{1}{2\pi} \frac{d \sigma}{d |t| \, d\xi \, d\cos\theta} 
		\bb{ 1 + s_T \, A_N(t, \xi, \cos\theta) \sin\phi_S },
\label{eq:cross-section}
\end{align}
where the unpolarized cross section is 
\beq[eq:unpol-cross-section]
	\frac{d \sigma}{d |t| \, d\xi \, d\cos\theta} 
	= 2 \pi \pp{ \alpha_e \alpha_s \frac{C_F}{N_c} }^2 \frac{1}{\xi^2 s^3} \Sigma_U(t, \xi, \cos\theta),
\eeq
with $\Sigma_U$ given by
\begin{align}
	&\Sigma_U^{\gamma} = (1 - \xi^2) \sum_{\alpha=\pm} \pp{ | \M_{\alpha}^{[\wt{H}]} |^2 + | \wt{\M}_{\alpha}^{[H]} |^2 }
	\nn\\
	& \hspace{2.5em}
		- \pp{ \xi^2 + \frac{t}{4m^2} } \sum_{\alpha=\pm} | \wt{\M}_{\alpha}^{[E]} |^2
		- \frac{\xi^2 t}{4m^2} \sum_{\alpha=\pm} | \M_{\alpha}^{[\wt{E}]} |^2 	\nn\\
	& \hspace{2.5em}
		- 2 \xi^2 \sum_{\alpha=\pm} \Re\pp{ \wt{\M}_{\alpha}^{[H]} \wt{\M}_{\alpha}^{[E] *}  
			+ \M_{\alpha}^{[\wt{H}]} \M_{\alpha}^{[\wt{E}]*} }
\label{eq:sigma-U-a}
\end{align}
for the diphoton production process,
with $m$ being the nucleon mass and the factorized helicity amplitudes $\M_\pm$ and $\wt{\M}_\pm$
as defined in \eq{eq:diphoton-convolution}, and
\begin{align}
	\Sigma_U^{\ell} 
	= &\; (1 - \xi^2) | \M_{0}^{[\wt{H}]} |^2
		- \frac{\xi^2 t}{4m^2} | \M_{0}^{[\wt{E}]} |^2 	\nn\\
	&\, - 2 \xi^2 \Re \pp{ \M_{0}^{[\wt{H}]} \M_{0}^{[\wt{E}]*} }
\label{eq:sigma-U-l}
\end{align}
resulting for the DY dilepton production process,
with the factorized helicity amplitude $\M_0$ as given in \eq{eq:dilepton-M0}.
$A_N$ in \eq{eq:cross-section} is the single nucleon transverse spin asymmetry (SSA)
given by
\begin{align}
	A_N^{\gamma} = &\,\frac{(1 + \xi)}{\Sigma_U^{\gamma}} \frac{\Delta_T}{m}
		\sum_{\alpha=\pm} \Im \bb{ \wt{\M}_{\alpha}^{[H]} \wt{\M}_{\alpha}^{[E] *}
			- \xi \M_{\alpha}^{[\wt{H}]} \M_{\alpha}^{[\wt{E}]*} 
		}
\label{eq:ssa-a}
\end{align}
for the diphoton process, and
\begin{align}
	A_N^{\ell} = &\,\frac{(1 + \xi)}{\Sigma_U^{\ell}} \frac{\Delta_T}{m} \, (-\xi)\,
		\Im \M_0^{[\wt{H}]} \M_0^{[\wt{E}]*}
\label{eq:ssa-l}
\end{align}
for the DY dilepton process.

The SSA arises from the interference of opposite target helicity states. 
Contrary to the SSAs in inclusive processes, with either collinear~\cite{Qiu:1991pp,Qiu:1998ia,Kanazawa:2000hz,Kanazawa:2000cx,Metz:2012ct}
or transverse-momentum-dependent factorizations~\cite{Sivers:1989cc,Sivers:1990fh,Collins:1992kk},
where the target helicity flip is achieved by the dynamics of partons,
the SSA here (of the exclusive process) is due to the transverse momentum shift in the diffractive subprocess,
which compensates the target helicity flip with a nonzero orbital angular momentum.
Hence the SSAs in \eqs{eq:ssa-a}{eq:ssa-l} are proportional to $\Delta_T$.
By parity invariance, only the $\bm{s}_T$ component perpendicular to the diffractive plane can play a role,
which explains the $\sin\phi_S$ modulation in \eq{eq:cross-section}.
The SSA can be measured by extracting the coefficient of this azimuthal modulation
or directly from the ``up-down'' asymmetry of $\bm{s}_T$ with respect to the diffractive plane,
\begin{align}
	&s_T \, A_N(t, \xi, \cos\theta) \nn\\
	&\hspace{1.5em}
	= \frac{\pi}{2} \cdot \frac{d\sigma(\bm{s}_T\cdot \hat{y}_{\rm lab} > 0) - d\sigma(\bm{s}_T\cdot \hat{y}_{\rm lab} < 0)}{
			d\sigma(\bm{s}_T\cdot \hat{y}_{\rm lab} > 0) + d\sigma(\bm{s}_T\cdot \hat{y}_{\rm lab} < 0)},
\end{align}
where $d\sigma(\bm{s}_T\cdot \hat{y}_{\rm lab} \gtrless 0)$ 
is the $\phi_S$-integrated cross section (still differential in $t$, $\xi$, and $\cos\theta$)
for the events with $\bm{s}_T$ aligning with or opposite to the direction of $\hat{y}_{\rm lab}$.

The single target transverse spin asymmetry has been studied~\cite{Ellinghaus:2005uc, Goloskokov:2008ib, Goloskokov:2013mba, Diehl:2007jy, Diehl:2005pc} and measured~\cite{HERMES:2009gyx, HERMES:2011bou, HERMES:2008abz, COMPASS:2012ngo, COMPASS:2013fsk, COMPASS:2016ium} for DVCS and DVMP. The analysis is performed mainly in the Breit frame, where the SSA exhibits an azimuthal correlation between the final-state photon and the target spin $\bm{s}_T$. For the mesoproduction processes here, in contrast, it is convenient to work in the two-stage picture, where the SSA only concerns the $\phi_S$ distribution in the lab frame.

Notably, all occurrences of the GPD $E$ or $\wt{E}$ are suppressed by $t$ and/or $\xi$,
which strongly limits the experimental capacity to measure them for moderate values.
In particular, in the unpolarized cross section, \eqs{eq:sigma-U-a}{eq:sigma-U-l}, 
they cannot be easily separated from the background caused by $H$ and $\wt{H}$. 
However, the SSA depends linearly on $E$ and $\wt{E}$ (modulo the $1/\Sigma_U$ factor in front).
Although its magnitude is suppressed by $\Delta_T$ and $\xi$, the SSA can give 
clear signals for extracting $E$ and $\wt{E}$
if it can be measured precisely enough.
With that said, the GPD $\wt{E}$ has a significant enhancement from the pion-pole contribution, as required by the sum rule.
Its contribution dominates in both the cross section and SSA.
As we will see, this strongly suppresses the sensitivity to GPD $H$ and $E$
but effectively provides enhancement to $\wt{H}$.

%================================================================
\section{Sensitivity to the $x$ dependence of transition GPD${\bf s}$}
\label{sec:x}
%================================================================

While the $\xi$ and $t$ dependence of GPDs directly corresponds to the diffractive kinematics
which can be measured in experiments, the $x$ dependence is only indirectly accessed through 
convolutions with hard coefficients like those in \eqs{eq:factorize}{eq:dilepton-factorize}.
It is only the hard scale $q_T$ (or, equivalently, $\theta$) that can help 
probe the hadron's partonic structure via the $x$ dependence of the GPDs.

One can, however, immediately notice from \eq{eq:dilepton-coefs} that the $\theta$ dependence
of the leading-order hard coefficient for the DY dilepton process completely factorizes from
the $x$ and $z$ dependence. Its amplitude simply reduces to the {\it moments} of the GPDs,
\beq[eq:GPD-moment]
	\wt{F}^{\pm}_0(\xi, t) \equiv \P \int_{-1}^1 dx \frac{\wt{F}^{\pm}(x, \xi, t)}{x - \xi},
\eeq 
with $\P$ referring to principle-value integration,
with a completely predictable $\sin\theta$ distribution 
decoupled from the $x$ dependence of the GPDs.
Here $\wt{F}^{\pm}(x, \xi, t) = \wt{F}(x, \xi, t) \pm \wt{F}(-x, \xi, t)$
is the charge-conjugation-even or -odd GPD component [see \eq{eq:charge-conjugation-relation}]
and we used the same moment notations as in \eq{eq:GPD-DA-moments}.

This simple factorization property also applies to many other processes, such as DVCS, DVMP, and TCS,
which yield only moment-type sensitivity at leading order. 
As first demonstrated in \refcite{Bertone:2021yyz}, there are infinite families of analytic functions,
called shadow GPDs, that give zero to the moments in \eq{eq:GPD-moment}
and cannot be distinguished from the ``real GPD'' by those processes alone.

In contrast, for the diphoton process, the hard coefficients in \eq{eq:diphoton-coefs} 
contain terms in which the $\theta$ dependence is entangled with the $x$ (and $z$) dependence,
resulting in a special convolution integral for the GPDs,
\begin{align}
	&I[\F^+; \xi, t, z, \theta] \nn\\
	&\hspace{2em}
	= \int_{-1}^{1}dx \frac{\F^+(x, \xi, t)}{x - x_p(\xi, z, \theta) + i \epsilon \, \sgn{\cos^2(\theta/2) - z} },
\label{eq:diphoton-special-int}
\end{align}
with a new pole for the $x$ integration,
\beq[eq:x-pole]
	x_p(\xi, z, \theta)
		= \xi \cdot \bb{ \frac{1 - z + \tan^2(\theta / 2) \, z}{1 - z - \tan^2(\theta / 2) \, z} }.
\eeq
The special integral appears in the two helicity amplitudes $\M_-$ and $\wt{\M}_-$ in \eq{eq:diphoton-convolution}, 
with $\F = \wt{F}$ and $F$, respectively, and 
the ``$-$'' subscripts referring to the two photons having aligned spins.
The new pole in \eq{eq:x-pole} scans through $[\xi, \infty) \cup (-\infty, -\xi]$ as $z$ goes from $0$ to $1$
and reflects the GPD $x$ dependence in the Dokshitzer-Gribov-Lipatov-Altarelli-Parisi (DGLAP) region. 
In comparison, the DY dilepton and other similar processes provide sensitivity only on the ridges $x = \pm \xi$.

\begin{figure}[htbp]
	\centering
	\includegraphics[trim={0 0 2.3mm 0}, clip, scale=0.43]{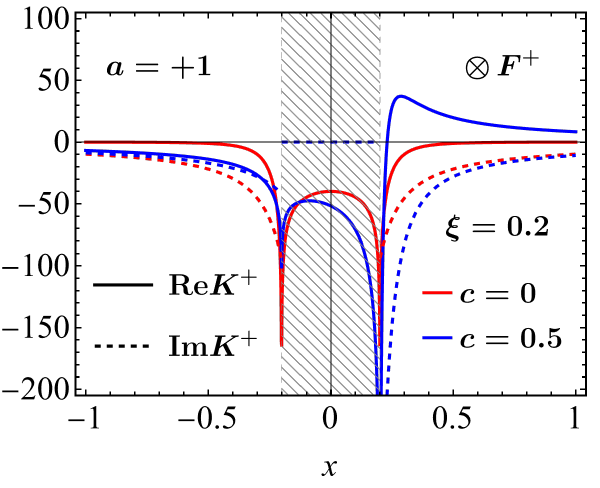}
	\includegraphics[trim={2mm 0 2mm 0}, clip, scale=0.43]{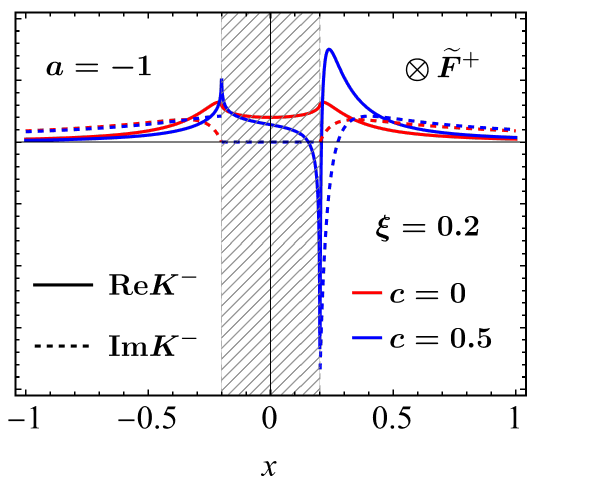}
	\caption{Kernels $K^{a}$ for the special $x$ integrals of GPDs $F$ and $\wt{F}$, respectively, 
	shown for $\xi = 0.2$ and two values of $c\equiv \cos\theta$. 
	Negative $c$ values can be seen from the symmetry $K^a(x, \xi, -c) = K^a(-x, \xi, c)$. 
	The solid and dashed lines refer to the real and imaginary parts, respectively.}
	\label{fig:kernel}
\end{figure}

The exact form of $x$ sensitivity by \eq{eq:diphoton-special-int} is further complicated by the 
integration with the DA and depends on the specific form of the latter. 
As a simple demonstration in this paper, we take $D^{[u\bar{d}]}_{\pi^+}(z) = -D^{[d\bar{u}]}_{\pi^-}(z) = (i f_{\pi} / 2) \phi(z)$,
with $f_{\pi} = 0.13~\GeV$ being the pion decay constant and $\phi(z) = 6z(1-z)$ the asymptotic form.
Then for the two amplitudes $\M_-$ and $\wt{\M}_-$ in \eqs{eq:diphoton-M-}{eq:diphoton-Mt-}, 
we define
\begin{align}\label{eq:diphoton-profiled}
	\bar{I}^a[\phi, \F^+; \xi, t, \theta]
	& \, = \int_0^1dz \, \frac{2\phi(z)}{z(1-z)} \cdot I[\F^+; \xi, t, z, \theta]		\nn\\
	& \hspace{1.5em}\times
		\bb{ \frac{\cos^2(\theta/2) - z}{\sin^2\theta} + \frac{a}{\cos^2(\theta/2) - z} }
		\nn\\
	& \, = \int_{-1}^{1}dx \, \F^+(x, \xi, t) \, K^a(x, \xi, \cos\theta),
\end{align}
where $a = +1$ for $\F = F$ and $-1$ for $\F = \wt{F}$, and the kernel $K^a$ is 
easily obtained with the asymptotic pion DA,
\begin{align}
	& 6^{-1} K^a(x, \xi, c)
	=  a \, L(x, \xi, c) 	\nn\\
	& \hspace{1.8em}
		+ \frac{\xi^2(1-c^2)}{(x - \xi c)^2} 
			\bb{ L(x, \xi, c) + \frac{1}{\xi(1-c^2)} + \frac{c (x - \xi c)}{\xi^2 (1-c^2)^2} }
	\nn\\
	& \hspace{1.8em} 
		- i \pi \, \frac{\theta(x^2 - \xi^2)}{2|x - \xi c|}
		\bb{ \frac{\xi^2 (1 - c^2)}{(x - \xi c)^2} + a },
\label{eq:diphoton-x-kernel}
\end{align}
where $c \equiv \cos\theta$ and 
\beq
	L(x, \xi, c) = \frac{1}{-2(x - \xi c)} \ln\abs{\frac{(1 - c)(x + \xi)}{(1 + c)(x - \xi)}}.
\eeq
As shown in \fig{fig:kernel}, 
the kernel $K^a$ exhibits a strong logarithmic behavior at $x\to \pm \xi$
and becomes flatter as $|x| \gg |\xi|$ or $|x| \ll |\xi|$.

%----------------------------------------------------------------
% Fig: kernels
%----------------------------------------------------------------
%----------------------------------------------------------------

\begin{figure}[htbp]
	\centering
	\includegraphics[trim={0 0 2.3mm 0}, clip, scale=0.43]{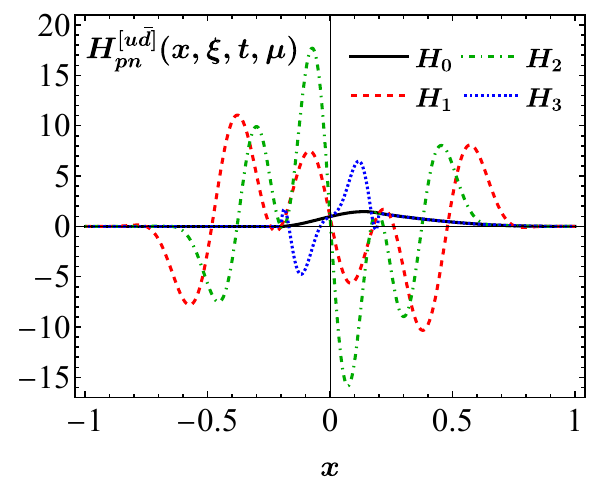}
	\includegraphics[trim={2mm 0 2mm 0}, clip, scale=0.43]{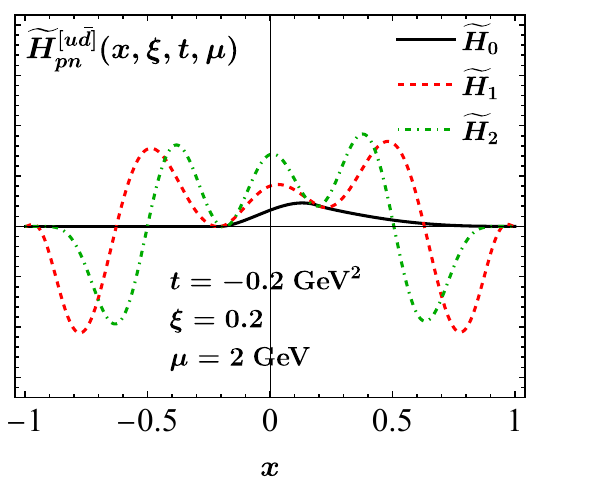}
	\caption{Transition GPD models for $H^{[u\bar{d}]}_{pn}$ and $\wt{H}^{[u\bar{d}]}_{pn}$ at $(t, \xi) = (-0.2~\GeV^2$, $0.2)$.
	The GPD $H_3$ is obtained from $H_0$ by adding a shadow $D$-term.}
	\label{fig:GPD-H}
\end{figure}

\begin{figure}[htbp]
	\centering
	\includegraphics[trim={0 0 2.3mm 0}, clip, scale=0.43]{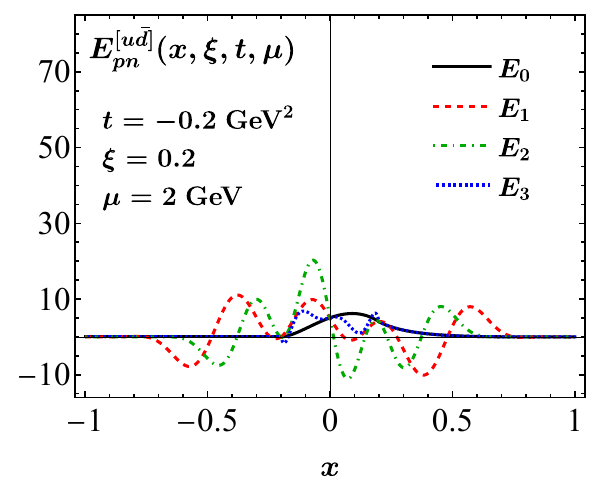}
	\includegraphics[trim={2mm 0 2mm 0}, clip, scale=0.43]{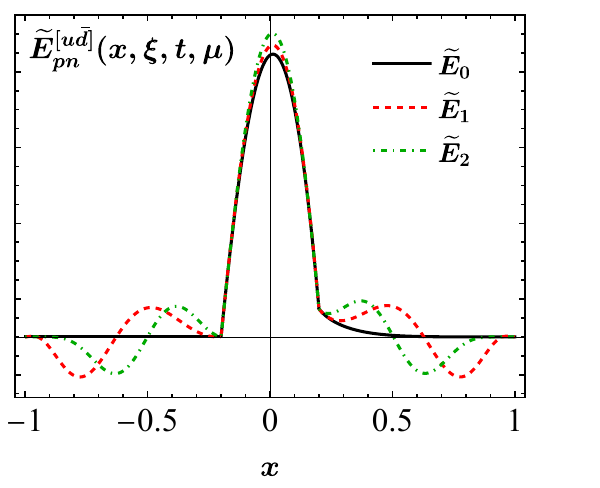}
	\caption{Same as \fig{fig:GPD-H}, but for the transition GPDs $E^{[u\bar{d}]}_{pn}$ and $\wt{E}^{[u\bar{d}]}_{pn}$.}
	\label{fig:GPD-E}
\end{figure}

Clearly, the diphoton process does have the capability of distinguishing between two GPD functions
that differ only by some shadow GPD by yielding different $\theta$ (or $q_T$) distributions,
while the DY dilepton process does not.
To demonstrate this, we construct the transition GPD models by adding two shadow GPDs,
$S_i(x, \xi)$ [$\wt{S}_i(x, \xi)$] with $i = 1$ or $2$, on $H^{[u\bar{d}]}_{pn}$ and $E^{[u\bar{d}]}_{pn}$
($\wt{H}^{[u\bar{d}]}_{pn}$ and $\wt{E}^{[u\bar{d}]}_{pn}$)
in the Goloskokov-Kroll (GK) model~\cite{Goloskokov:2005sd, Goloskokov:2007nt, Goloskokov:2009ia, Kroll:2012sm},
as in \refcite{Qiu:2023mrm}.
The specific forms of the shadow GPDs used in this paper are collected in Appendix \ref{app:sGPD}.
The GPDs $H$ and $E$ also allow a $D$-term that lives only in the Efremov-Radyushkin-Brodsky-Lepage (ERBL) region,
for which we may similarly construct a shadow $D$-term.
The resultant GPDs, denoted as $(H_i, E_i, \wt{H}_i, \wt{E}_i)$,
with $i = 0$ referring to the GK model, are shown in Figs.~\ref{fig:GPD-H} and \ref{fig:GPD-E},
where one can notice the strong pion-pole enhancement in the ERBL region of the GPD $\wt{E}$.

\begin{figure}[htbp]
	\centering
	\includegraphics[trim={0 0 1mm 0}, clip, scale=0.42]{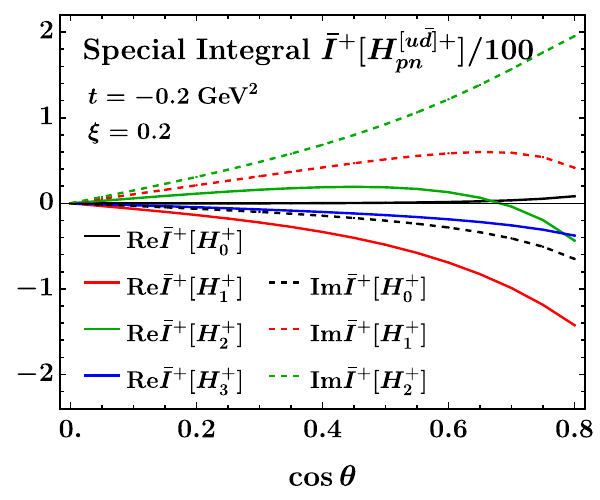}
	\includegraphics[trim={2mm 0 1mm 0}, clip, scale=0.42]{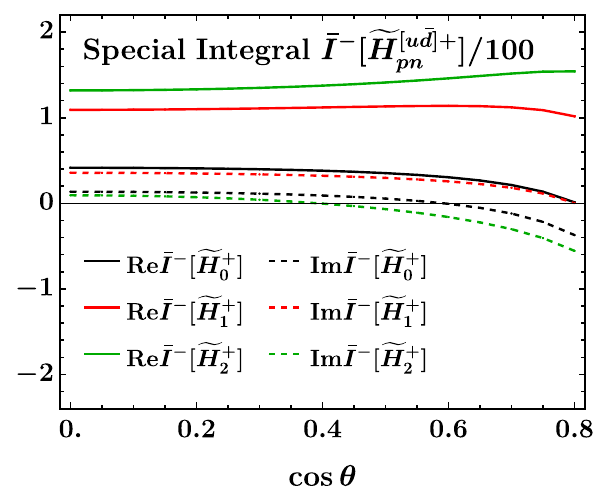}
	\caption{Special integrals $\bar{I}^{\pm}$ for the GPDs $H^{[u\bar{d}] +}_{pn}$ and $\wt{H}^{[u\bar{d}] +}_{pn}$ 
	evaluated at $(t, \xi) = (-0.2~\GeV^2, 0.2)$ for different GPD models labeled by the subscripts $0$--$3$.
	The solid and dashed lines refer to the real and imaginary parts, respectively.
	The shadow $D$-term in $H^{[u\bar{d}] +}_{pn, 3}$ does not contribute to the imaginary part,
	so the corresponding curve coincides with the dashed black line and is not shown.}
	\label{fig:spec-int-H}
\end{figure}

\begin{figure}[htbp]
	\centering
	\includegraphics[trim={0 0 1mm 0}, clip, scale=0.42]{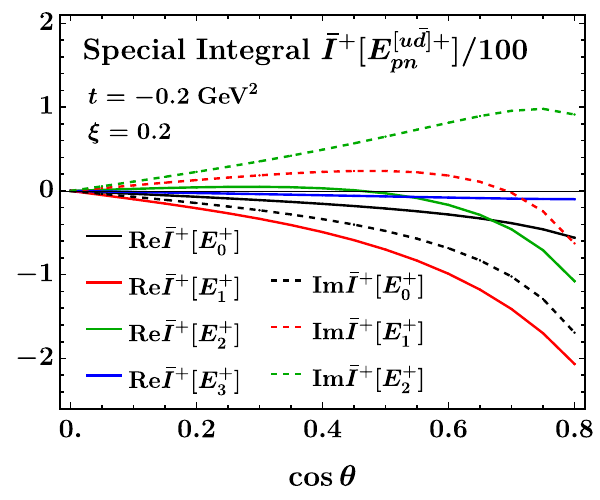}
	\includegraphics[trim={2mm 0 1mm 0}, clip, scale=0.42]{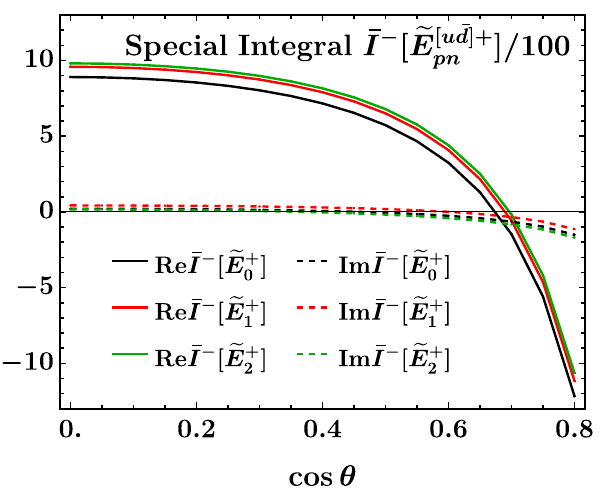}
	\caption{Same as \fig{fig:spec-int-H}, but for the GPD $E^{[u\bar{d}] +}_{pn}$ and $\wt{E}^{[u\bar{d}] +}_{pn}$.}
	\label{fig:spec-int-E}
\end{figure}

Evaluating the integral $\bar{I}^{\pm}$ for these GPD models 
gives different $\cos\theta$ distributions, as shown in Figs.~\ref{fig:spec-int-H} and \ref{fig:spec-int-E}.
$\bar{I}^+$ ($\bar{I}^-$) evaluated on $H^{[u\bar{d}] +}_{pn}$ or $E^{[u\bar{d}] +}_{pn}$ 
($\wt{H}^{[u\bar{d}] +}_{pn}$ or $\wt{E}^{[u\bar{d}] +}_{pn}$)   
gives an odd (even) distribution of $\cos\theta$, so it is shown only for the positive $\cos\theta$.
This property also strongly suppresses the value of $\bar{I}^+$ in the central rapidity region,
so the value of $\bar{I}^-$ is generally larger than $\bar{I}^+$.
We note that the shadow GPDs give substantial contributions to the special integrals,
altering both the magnitudes and the shapes of the GK model results.
The real and imaginary parts are both important in $\bar{I}^+$, 
whereas the real part dominates in $\bar{I}^-$,
which agrees with the kernel behavior in \fig{fig:kernel}.

%----------------------------------------------------------------
% Fig: GPDs H
%----------------------------------------------------------------
%----------------------------------------------------------------

%----------------------------------------------------------------
% Fig: GPDs E
%----------------------------------------------------------------
%----------------------------------------------------------------

%----------------------------------------------------------------
% Fig: special integral H
%----------------------------------------------------------------
%----------------------------------------------------------------

%----------------------------------------------------------------
% Fig: special integral E
%----------------------------------------------------------------
%----------------------------------------------------------------

%================================================================
\section{Numerical study for J-PARC and AMBER energies}
\label{sec:numeric}
%================================================================

Charged pion beams can be accessed in fixed target experiments 
at J-PARC~\cite{Aoki:2021cqa} and AMBER at CERN~\cite{Adams:2018pwt} facilities as secondary beams,
with wide energy coverage.
It is not the purpose of this paper to perform a realistic experimental simulation;
rather, we explore how the diphoton process can complement the DY dilepton one to
help disentangle transition GPDs from shadow GPDs.
At a benchmark study, we choose the two pion beam energies, $E_{\pi} = 20$ and $100~\GeV$, 
for J-PARC and AMBER, respectively,
and examine only the nucleon transitions.

\begin{figure}[htbp]
	\centering
	\includegraphics[trim={0 0 1mm 0}, clip, scale=0.37]{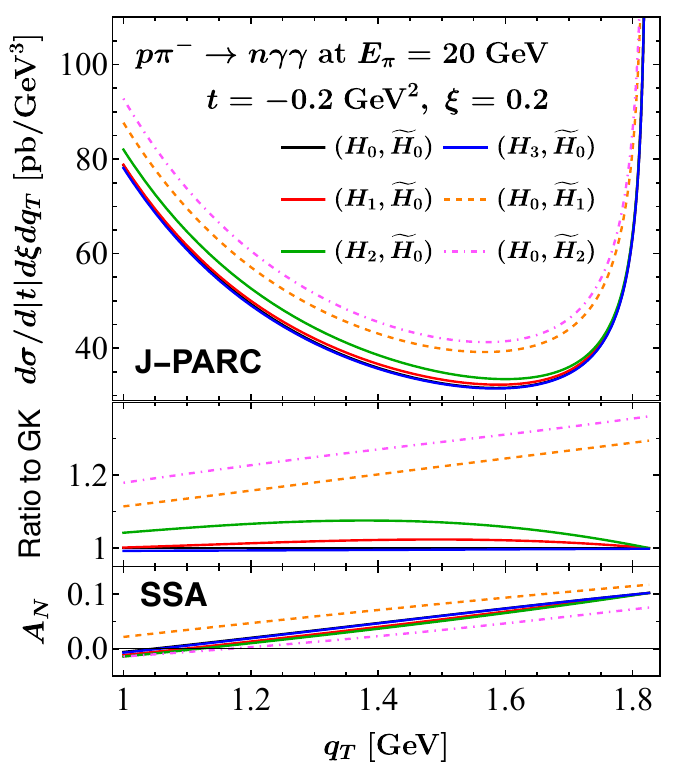}
	\includegraphics[trim={0 0 1mm 0}, clip, scale=0.37]{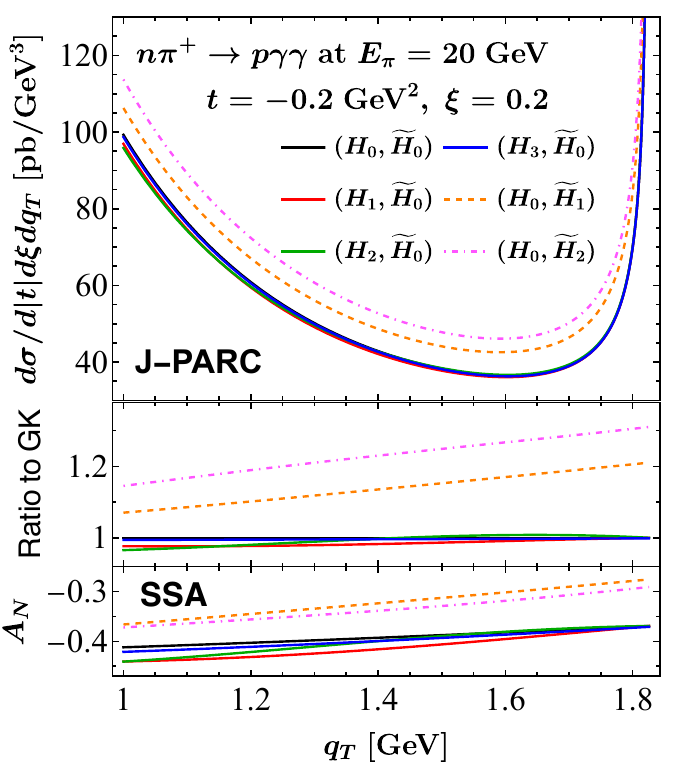}
	\caption{Unpolarized $q_T$ distributions and SSAs at $E_{\pi} = 20~\GeV$ and $(t, \xi) = (-0.2~\GeV^2, 0.2)$
	evaluated for different GPD models obtained by varying $H$ and $\wt{H}$.
	Left panel: $\pi^-p$ reaction. Right panel: $\pi^+n$ reaction.}
	\label{fig:qt-20}
\end{figure}

\begin{figure}[htbp]
	\centering
	\includegraphics[trim={0 0 1mm 0}, clip, scale=0.37]{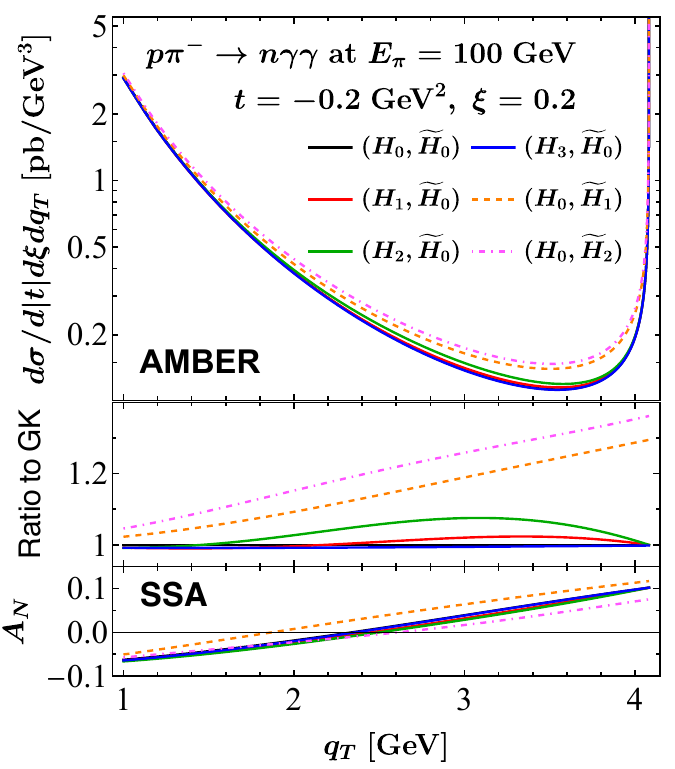}
	\includegraphics[trim={0 0 1mm 0}, clip, scale=0.37]{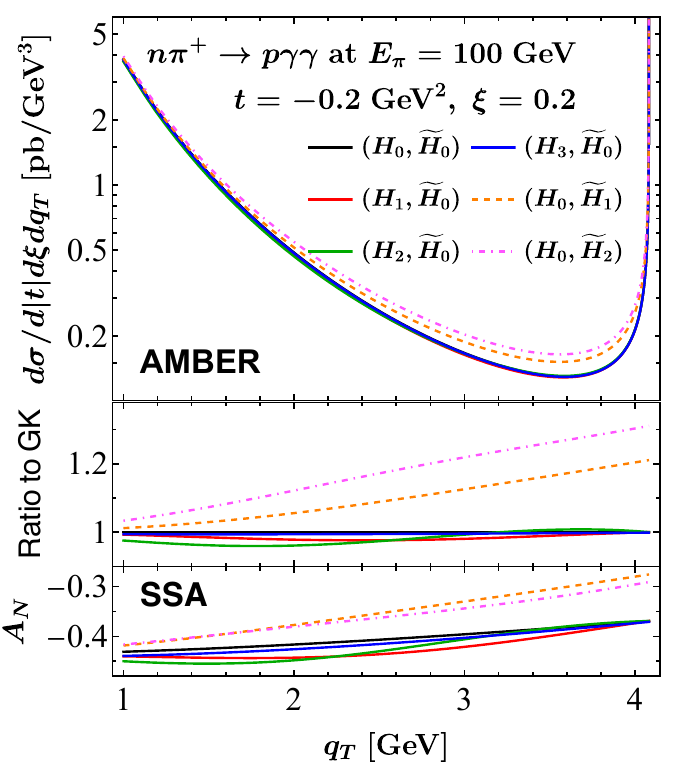}
	\caption{Same as \fig{fig:qt-20}, but for $E_{\pi} = 100~\GeV$. 
	The vertical scales of the upper panels are logarithmic.}
	\label{fig:qt-100}
\end{figure}

\begin{figure}[htbp]
	\centering
	\includegraphics[trim={0 0 1mm 0}, clip, scale=0.37]{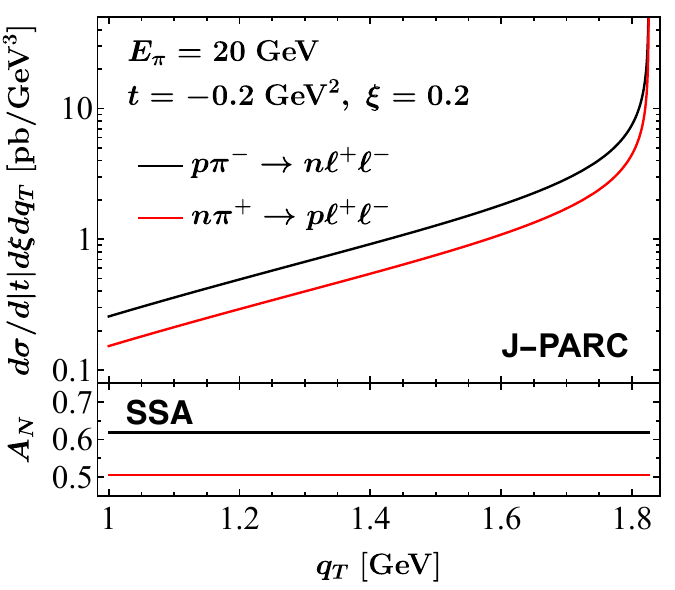}
	\includegraphics[trim={0 0 1mm 0}, clip, scale=0.37]{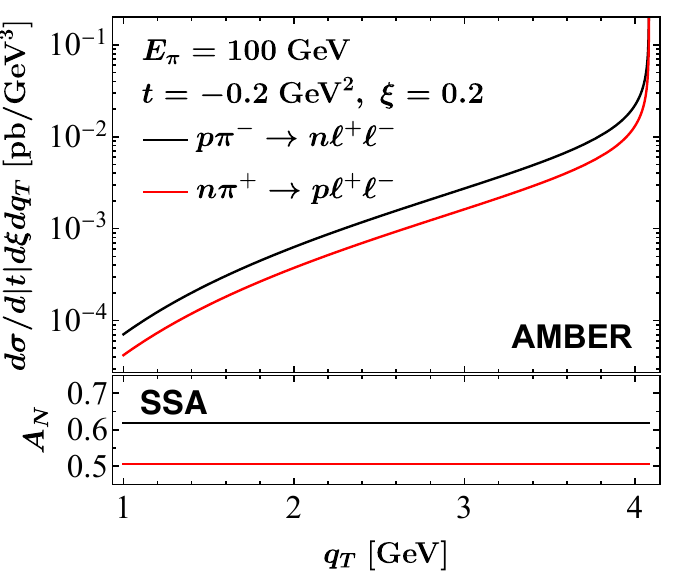}
	\caption{Unpolarized $q_T$ distributions and SSAs for the DY dilepton process.
	The vertical scales of the upper panels are logarithmic.}
	\label{fig:dilepton-qt}
\end{figure}

Evaluating the unpolarized cross sections in \eq{eq:unpol-cross-section} and the SSAs in \eq{eq:ssa-a} 
for the GPD models in Figs.~\ref{fig:GPD-H} and~\ref{fig:GPD-E} 
yields different distributions of $q_T$, as shown in Figs.~\ref{fig:qt-20} and~\ref{fig:qt-100} 
at $(t, \xi)$ $=$ $(-0.2~\GeV^2, 0.2)$ for the $p\pi^-$ and $n\pi^+$ collisions 
at $E_{\pi} = 20$ and $100~\GeV$, respectively.
We require $q_T \geq 1~\GeV$ to constrain the power correction from $\sqrt{|t|} / q_T$.
The $q_T$ peaks of cross sections on the right are due to Jacobian effects.
In each choice of the GPD models, we vary $H^{[u\bar{d}]}_{pn}$ or $\wt{H}^{[u\bar{d}]}_{pn}$ one at a time
and fix $E^{[u\bar{d}]}_{pn}$ and $\wt{E}^{[u\bar{d}]}_{pn}$ to be of the GK model; 
varying the latter makes minimal differences due to the kinematic suppression factors in \eq{eq:sigma-U-a}.
We note that varying the GPDs yields not only different magnitudes for both production rates and SSAs
but also different shapes in the $q_T$ distributions.
Thus, in experimental analyses with binned $q_T$ observables, 
each bin offers an independent constraint on the GPDs, 
in contrast to the DY dilepton process, where the $q_T$ shape is universal (at leading order), 
as shown in \fig{fig:dilepton-qt}.
 
The AMBER energy gives much wider $q_T$ coverage, but much smaller (differential) production rates,
and leads to similar sensitivity in terms of the rate distribution ratios.
In all cases, we notice that the shadow $D$-term added to $H^{[u\bar{d}]}_{pn}$ does not make 
as much an impact as it does for the photoproduction process in \refcite{Qiu:2023mrm},
as a result of the special pole $x_p$ in \eq{eq:x-pole} lying only in the DGLAP region.

The difference between the $p\pi^-$ and $n\pi^+$ collisions is caused by the isospin-breaking effect of QED interactions, 
namely, the $(e_1^2 - e_2^2)$ terms in \eq{eq:diphoton-convolution}. 
Numerically, the diagonal values $\F(\xi, \xi, t)$ of transition GPDs $\F$ dominate over the their moments $\F_0(\xi, t)$
for a wide range of $\xi$ at a given $t$, although sea quark components do not contribute to transition GPDs.
The $(e_1^2 - e_2^2)$ terms are therefore non-negligible and cause significant differences in the imaginary parts of the amplitudes
in \eq{eq:diphoton-convolution} between the $p\pi^-$ and $n\pi^+$ collisions.
This induces the differences of cross sections and SSAs between the two channels.

%----------------------------------------------------------------
% Fig: cross section at 20GeV
%----------------------------------------------------------------
%----------------------------------------------------------------

%----------------------------------------------------------------
% Fig: cross section at 100GeV
%----------------------------------------------------------------
%----------------------------------------------------------------

%----------------------------------------------------------------
% Fig: dilepton 
%----------------------------------------------------------------
%----------------------------------------------------------------

The GPDs $H^{[u\bar{d}]}_{pn}$ and $E^{[u\bar{d}]}_{pn}$ make much smaller contributions to the cross sections and SSAs
than $\wt{H}^{[u\bar{d}]}_{pn}$ and $\wt{E}^{[u\bar{d}]}_{pn}$ because
(1) when making the flavor transition GPDs, there is a cancellation (enhancement) between $u$ and $d$ quark GPDs for the former (latter) 
	in the GK model employed here;
(2) aside from the $(e_1^2 - e_2^2)$ terms in \eq{eq:diphoton-convolution}, it is the $C$-even GPD combinations that contribute to the amplitudes,
which gives a further cancellation (enhancement) for the former (latter) [see \eq{eq:charge-conjugation-relation}]; 
and
(3) most importantly, the pion-pole term in $\wt{E}^{[u\bar{d}]}_{pn}$ severely leverages the contributions from $\wt{H}^{[u\bar{d}]}_{pn}$,
	overshadowing the effects from $H^{[u\bar{d}]}_{pn}$ and $E^{[u\bar{d}]}_{pn}$.
Hence, the $q_T$ distributions are more sensitive to the GPD $\wt{H}^{[u\bar{d}]}_{pn}$ than $H^{[u\bar{d}]}_{pn}$ for both the cross sections and SSAs.
However, one would still be able to probe the GPD $H^{[u\bar{d}]}_{pn}$ near $|x| \sim \xi$, 
which is why the shadow GPDs $S_i(x, \xi)$ for $H^{[u\bar{d}]}_{pn}$ have been constructed to be much more dramatic 
in this region than the $\wt{S}_i(x, \xi)$ for $\wt{H}^{[u\bar{d}]}_{pn}$.
If we remove or decrease the pion-pole contribution from $\wt{E}^{[u\bar{d}]}_{pn}$, 
the sensitivity to $H^{[u\bar{d}]}_{pn}$ would become more manifest.

Despite the kinematic suppression in \eq{eq:ssa-a}, we get significant SSAs, especially for the $\pi^+n$ channel,
which is dominated by the large pion-pole enhancement in the GPD $\wt{E}^{[u\bar{d}]}_{pn}$.
Hence, the SSAs can also serve as useful probes of GPDs, 
particularly of $\wt{H}^{[u\bar{d}]}_{pn}$, 
since it is the one that interferes with $\wt{E}^{[u\bar{d}]}_{pn}$ in \eq{eq:ssa-a},
although it has one extra suppression factor of $\xi$,
as shown in Figs.~\ref{fig:qt-20} and \ref{fig:qt-100}.
For the same reason, varying the GPDs $E^{[u\bar{d}]}_{pn}$ and $\wt{E}^{[u\bar{d}]}_{pn}$ 
by the same shadow GPDs does not yield appreciable effects.
Nevertheless, by simply scaling them with some factor, the SSAs are proportionally changed. 
Measuring the SSA gives sensitive probes to both the magnitudes and the signs of the GPDs $E^{[u\bar{d}]}_{pn}$ and $\wt{E}^{[u\bar{d}]}_{pn}$.

In comparison, for the DY dilepton process, the shadow GPDs are constructed to give null impacts.
The above GPD models thus give the same prediction for both the unpolarized cross section in \eq{eq:sigma-U-l}
and the SSA in \eq{eq:ssa-l}, shown in \fig{fig:dilepton-qt} for the J-PARC and AMBER energies, respectively.
At leading power, the virtual photon decaying into the lepton pair is purely longitudinally polarized,
giving a $\sin^2\theta$ angular distribution [\eq{eq:dilepton-M0}], so the cross section is finite at small $q_T$.
This feature contrasts with the diphoton process, whose production rate is enhanced in the forward region, with a much larger overall rate as well.
Interestingly, the DY dilepton process produces a large SSA.
It does not depend on $q_T$, and the energy dependence cancels in the ratio of \eq{eq:ssa-l}.
This measurement could set a useful constraint on the overall magnitude of 
GPD $\wt{E}^{[u\bar{d}]}_{pn}$, especially on the pion-pole peak in the GK model,
but has no capability of probing its $x$ dependence.

%================================================================
\section{Conclusion}
\label{sec:conclusion}
%================================================================

While GPDs encode valuable information on hadrons' emergent properties and tomographic structures,
it is a challenging task to fully extract such information out of experiments, especially the $x$ dependence.
In this paper, we compared two single diffractive hard exclusive mesoproduction processes, 
the DY dilepton production and diphoton production, 
which can be studied at hadron experiment facilities like J-PARC and AMBER at CERN using secondary pion beams.
At leading power, the scattering amplitudes of both processes can be factorized into transition GPDs
and provide constraints on the nucleon GPDs through isospin symmetry, 
complementary to the electroproduction and photoproduction processes.

We showed that both the unpolarized production rate and the target spin asymmetry can help probe the transition GPDs
and clearly demonstrated that, while the DY dilepton process only constrains the polarized GPDs through simple moments,
the diphoton process can provide enhanced $x$ sensitivity from transverse momentum ($q_T$) distributions
of the final-state particles that are entangled with the $x$ dependence of the GPDs.
This enhanced $x$ sensitivity shares the same physics origin as that appearing in photoproduction of photon-meson pair studied in \refcite{Qiu:2023mrm}
but complements it by being more sensitive to the DGLAP region near $x = \pm\xi$.
Both J-PARC and AMBER at CERN give similar sensitivity; 
the former produces larger rates, but the latter covers a wider kinematical range.

Given the fact that we need to solve an inverse problem to extract GPDs from experimental data, 
we would not expect one or two simple processes to be sufficient to fully map out the $x$ dependence of the GPDs.
We therefore advocate that global analyses should be performed by combining data from various processes that 
are sensitive to the universal GPDs, especially those with enhanced $x$ sensitivity.  
Only in this way would it be possible to obtain the full tomographic images of the hadrons.

%================================================================
\section*{Acknowledgements}
%================================================================
We thank W.-C.~Chang, P.~Kroll, J.-C.~Peng, B.~Pire, M.~G.~Santiago, M.~Strikman, and N.~Tomida for the helpful discussions and communications. 
This work is supported in part by U.S. Department of Energy
(DOE) Contract No.~DE-AC05-06OR23177, under which
Jefferson Science Associates, LLC, operates Jefferson Lab.
The work of Z.Y. at MSU is partially supported by the 
U.S.~National Science Foundation under Grant No.~PHY-2013791 
and the fund from the Wu-Ki Tung endowed chair in particle physics.

\newpage
%%%%%%%%%%%%%%%%%%%%%%%%%%%%%%%%%%%%%%%%%%%%%%%%%%%%%%%%
\appendix
%%%%%%%%%%%%%%%%%%%%%%%%%%%%%%%%%%%%%%%%%%%%%%%%%%%%%%%%

\begin{widetext}
%================================================================
\section{HARD COEFFICIENTS OF THE DIPHOTON PRODUCTION PROCESS IN EQ.~\eqref{eq:diphoton}}
\label{app:diphoton}
%================================================================

As discussed in detail in \refs{Qiu:2023mrm, Qiu:2022pla, Qiu:2022bpq},
the hard coefficients in \eq{eq:factorize} can be conveniently calculated in the ERBL region 
for the hard scattering subprocess
by introducing the change of variables,
$(z_1, z_2) = ((x + \xi) / (2\xi), z)$, 
and carefully keeping track of all the $i\epsilon$ prescriptions.
This is directly analogous to the corresponding $2\to2$ meson annihilation process
and is well suited to the two-stage SDHEP picture.
It makes manifest the symmetry property between $z_1$ and $z_2$,
although they have different ranges, $z_1 \in [(-1 + \xi) / (2\xi), (1 + \xi) / (2\xi)]$ and $z_2 \in [0, 1]$. 
By parity property, the diphoton helicity amplitudes can be reduced to four independent hard coefficients,
\begin{align}
	C^{\gamma}_{++} = C^{\gamma}_{--} = \frac{\N}{\hat{s}} \, C^{\gamma}_+, &\quad
	C^{\gamma}_{+-} = C^{\gamma}_{-+} = \frac{\N}{\hat{s}} \, C^{\gamma}_-,	\nn\\
	\wt{C}^{\gamma}_{++} = - \wt{C}^{\gamma}_{--} = \frac{\N}{\hat{s}} \, \wt{C}^{\gamma}_+, &\quad
	\wt{C}^{\gamma}_{+-} = - \wt{C}^{\gamma}_{-+} = \frac{\N}{\hat{s}} \, \wt{C}^{\gamma}_-,
\label{eq:diphoton-helicity-structure}
\end{align}
where $\N = 2i e^2 g^2 C_F / N_c$ and the four independent hard coefficients are 
\bse\label{eq:diphoton-coefs}\begin{align}
%C+
	&2\xi C^{\gamma}_{+} =
		\pp{e_1-e_2}^2 \frac{2}{\sin^2\theta} \cdot
			\P\frac{z_1 z_2 + (1-z_1)\, (1-z_2)}{ z_1 z_2 (1-z_1) (1-z_2)}
		\nn\\
		& \hspace{3.5em} 
		+ \frac{2 i \pi}{\sin^2\theta} \cdot
			\bb{ (e_1^2-e_2^2)
				\pp{ \frac{\delta(z_1)}{z_2} - \frac{\delta(1-z_1)}{1-z_2} }
				- 2 e_1 e_2 
				\pp{ \frac{\delta(z_1)}{z_2} + \frac{\delta(1-z_1)}{1-z_2} }
			},
\label{eq:diphoton-C+}		\\
% C-
	&2\xi C^{\gamma}_{-}  = 
 		\pp{e_1-e_2}^2 \frac{2}{ \sin^2\theta } 
			\cdot \P \frac{z_1 + z_2 - 2z_1 z_2}{z_1 z_2(1-z_1)(1-z_2)} 	
		+ (e_1^2 - e_2^2) \cdot \P \frac{z_1 - z_2}{z_1 z_2(1-z_1)(1-z_2)} 
	\nn\\
	& \hspace{3.5em} 
		+ \P \frac{2 e_1 e_2}{z_1 z_2 (1-z_1) (1-z_2) } \cdot
		\frac{ \pp{ z_1(1-z_1) + z_2(1-z_2) } \pp{ z_1 z_2 + (1-z_1)(1-z_2) } }{
			\pp{ 2 z_1 z_2 + (1 - \cos\theta) (1-z_1-z_2) } 
			\pp{ 2 z_1 z_2 + (1 + \cos\theta) (1 - z_1 - z_2) }}
	\nn\\
	& \hspace{3.5em} 
	 	- i\pi \cdot \cc{ 
			\pp{e_1-e_2}^2 \pp{\frac{\delta (1-z_1)}{z_2} + \frac{\delta (z_1)}{1-z_2}}
			- (e_1^2 - e_2^2) \frac{2}{\sin^2\theta} \pp{ \frac{\delta (z_1)}{1-z_2} - \frac{\delta(1-z_1)}{z_2} }
			\right.
			\nn\\
			& \hspace{6em}
			+ 2 e_1 e_2 \bb{
				\frac{1 + \sin^2\theta}{\sin^2\theta}
					\pp{ \frac{\delta(1-z_1)}{z_2} + \frac{\delta (z_1)}{1-z_2} }
				\right.\nn\\
				& \hspace{10em} \left.
				+ \frac{1}{\sin^2\theta} 
					\pp{ \sgn{\cos^2(\theta/2) - z_2} \, \delta(z_1 - \rho(z_2)) \,
							\bigg( \frac{\cos\theta}{z_2(1-z_2)} - \frac{1}{\cos^2(\theta/2) - z_2} \bigg)
						 \right. \right.\nn\\
						& \hspace{14em} \left. \left. \left. 
							+ \, \sgn{z_2 - \sin^2(\theta/2)} \, \delta(z_1 - \wt{\rho}(z_2)) \,
							\bigg( \frac{\cos\theta}{z_2(1-z_2)} - \frac{1}{z_2 - \sin^2(\theta/2)} \bigg)
					}
			}
		},
\label{eq:diphoton-C-}		\\
% Ct+
	&2\xi\Ct^{\gamma}_+ = 
		\pp{e_1-e_2}^2 \frac{ -2 }{ \sin^2\theta } \cdot
 			\P\frac{1 - z_1-z_2}{z_1\, z_2\, (1-z_1)(1-z_2)}
		\nn\\
		& \hspace{3.5em} 
		-\frac{2 \pi i}{ \sin^2\theta }  \cdot \bb{
			(e_1^2 - e_2^2) \pp{ \frac{\delta(1-z_1)}{ 1 - z_2} + \frac{\delta (z_1)}{z_2} }
			- 2 e_1 e_2  \pp{ \frac{\delta (z_1)}{z_2}  -  \frac{\delta(1-z_1)}{ 1 - z_2} }
		},
\label{eq:diphoton-Ct+}	\\
% Ct-
	&2\xi\Ct^{\gamma}_- =
 		\pp{e_1-e_2}^2 \frac{2\cos\theta}{ \sin^2\theta } 
			\cdot \P \frac{z_1 - z_2}{z_1 z_2 (1-z_1)(1-z_2)}
		\nn\\
	& \hspace{3.5em}
		+ \P \frac{2 e_1 e_2 \, \cos\theta}{z_1 z_2 (1-z_1) (1-z_2) } \cdot
			\frac{ (z_1 - z_2) (1 - z_1 - z_2)^2 }{ 
				\pp{ 2 z_1 z_2 + (1 - \cos\theta) (1-z_1-z_2) } 
				\pp{ 2 z_1 z_2 + (1 + \cos\theta) (1-z_1-z_2) } 
			} \nn\\
	&\hspace{3.5em} 
			- \frac{2 \pi i }{\sin^2\theta} \cdot
			\cc{ 
				(e_1^2 - e_2^2) \cos\theta 
					\pp{ \frac{\delta(1-z_1)}{z_2} + \frac{\delta (z_1)}{1-z_2} }
				- e_1 e_2 
				\bb{ 
					\cos\theta \pp{ \frac{\delta (z_1)}{1-z_2} - \frac{\delta(1-z_1)}{z_2} }
					\right. \right.\nn\\
					&\hspace{8.5em} \left. \left.
					+ \frac{z_1 - z_2}{z_2 (1 - z_2)}
						\pp{ 
							\sgn{ \cos^2(\theta/2) - z_2} \delta(z_1-\rho(z_2)) - \sgn{\sin^2(\theta/2) - z_2} \delta(z_1-\wt{\rho}(z_2)) 
						}
				}
			},	
\label{eq:diphoton-Ct-}
\end{align}
\ese
where $\P$ indicates that the hard coefficients should be understood in the principle-value sense
when convoluted with the GPD and DA, and 
\begin{align}\label{eq:special-z1-pole}
	\rho(z_2) = \frac{\cos^2(\theta/2) \, (1 - z_2) }{\cos^2(\theta/2) -  z_2},
	\quad
	\wt{\rho}(z_2) = \frac{\sin^2(\theta/2) \, (1 - z_2) }{\sin^2(\theta/2) -  z_2}
\end{align}
are the special poles of $z_1$,
with small imaginary parts by the $i\epsilon$ prescription, respectively,
\beq
	i \epsilon \, \sgn{ z_2 - \cos^2(\theta/2)},
	\quad
	i \epsilon \, \sgn{ z_2 - \sin^2(\theta/2)}.
\eeq
We have expressed the hard coefficients in the general flavor case,
where the two parton lines $q$ and $q'$ carry the electric charges $e_1$ and $e_2$, respectively,
with $e_u = 2/3$ and $e_d = -1/3$.

The convolution of the hard coefficients in \eq{eq:diphoton-coefs} with the GPDs can be simplified 
by using the DA symmetry property $D(z_2) = D(1 - z_2)$,
\bse\label{eq:diphoton-convolution}\begin{align}
% M+
	\M_+^{[\wt{F}]} = &
		- \frac{2 D_0}{\sin^2\theta} \cc{
			\pp{e_1-e_2}^2 \, \wt{F}^+_0(\xi, t)
			+ i\pi \bb{
				(e_1^2-e_2^2)\, \wt{F}^-(\xi, \xi, t)
				+ 2 e_1 e_2 \, \wt{F}^+(\xi, \xi, t)
			}
		},
\label{eq:diphoton-M+}	\\
% M-
	\M_-^{[\wt{F}]}  = &
		- (e_1 - e_2)^2 \, D_0 \bb{ \frac{2}{\sin^2\theta} \, \wt{F}^+_0(\xi, t) 
				+ i \pi \wt{F}^+(\xi, \xi, t)
		}	
		- (e_1^2-e_2^2) \, D_0 \bb{ \wt{F}^-_0(\xi, t) + \frac{2i\pi}{\sin^2\theta} \wt{F}^-(\xi, \xi, t) }
		\nn\\
	& + e_1 e_2	\cc{
		\int_0^1dz \frac{2D(z)}{z(1-z)}
			\bb{ \frac{\cos^2(\theta/2) - z}{\sin^2\theta} - \frac{1}{\cos^2(\theta/2) - z} } 
			\cdot I[\wt{F}^+; \xi, t, z, \theta]
		\right.\nn\\
		&\hspace{12em} \left.
		- \frac{2 D_0}{\sin^2\theta} \cdot 
		\bb{ \wt{F}^+_0(\xi, t)
	 		+ i \pi \, (1 + \sin^2\theta) \, \wt{F}^+(\xi, \xi, t)
		}
	},
\label{eq:diphoton-M-} \\
% Mt+
	\wt{\M}_+^{[F]} = &
		-\frac{2D_0}{\sin^2\theta} \cc{
			(e_1 - e_2)^2 \, F^+_0(\xi, t)
			+ i \pi \bb{
				(e_1^2-e_2^2)\, F^-(\xi, \xi, t)
				+ 2 e_1 e_2 \, F^+(\xi, \xi, t)
			}
		},
\label{eq:diphoton-Mt+}	\\
% Mt-
	\wt{\M}_-^{[F]} = &
		- \frac{2\cos\theta}{\sin^2\theta} \cdot D_0 \bb{
			(e_1 - e_2)^2 \, F^+_0(\xi, t)
			+ i \pi \, (e_1^2-e_2^2)\, F^-(\xi, \xi, t)
		}
		\nn\\
		&+ e_1 e_2 \cc{
			\int_0^1dz \frac{2D(z)}{z(1-z)}
				\bb{ \frac{\cos^2(\theta/2) - z}{\sin^2\theta} + \frac{1}{\cos^2(\theta/2) - z} } \cdot
				I[F^+; \xi, t, z, \theta]
				\right.\nn\\
				&\hspace{12em} \left.
				- \frac{2 \cos\theta \, D_0}{\sin^2\theta} \cdot 
				\bb{ 
					F^+_0(\xi, t) + i \pi \, F^+(\xi, \xi, t)
				}
		},
\label{eq:diphoton-Mt-}
\end{align}\ese
where we use $F$ to denote either GPD $H$ or GPD $E$ and introduce the shorthand notations
\begin{align}
	\M_{\alpha}^{[\wt{F}]} \equiv \int_{-1}^1 dx \int_0^1 dz \, 
		\wt{F}(x, \xi, t) \, D(z) \, C_{\alpha}(x, \xi; z; \theta),
	\quad
	\wt{\M}_{\alpha}^{[F]}  \equiv \int_{-1}^1 dx \int_0^1 dz \, 
		F(x, \xi, t) \, D(z) \, \wt{C}_{\alpha}(x, \xi; z; \theta),
\label{eq:M-conv-short}
\end{align}
with $\alpha$ denoting any helicity index.
We have defined the “zeroth moments” of the DA and GPDs as
\beq[eq:GPD-DA-moments]
	D_0 \equiv \int_0^1 \frac{dz \, D(z)}{z}, 
	\quad
	\F_0(\xi, t) \equiv \P \int_{-1}^1 \frac{dx \, \F(x, \xi, t)}{x - \xi},
\eeq
and the special GPD integral as
\beq[eq:diphoton-special-int-app]
	I[\F; \xi, t, z, \theta] 
		\equiv 
		\int_{-1}^{1}dx \frac{\F(x, \xi, t)}{x - x_p(\xi, z, \theta) + i \epsilon \, \sgn{\cos^2(\theta/2) - z} },
\eeq
where $\F$ can take any GPD function ($F^{\pm}$, $\wt{F}^{\pm}$, $H^{\pm}$, etc.) 
whose superscripts ``$\pm$'' refer to the charge-conjugation parity following \refcite{Diehl:2003ny},
\begin{align}
	&F^{\pm}(x, \xi, t) = F(x, \xi, t) \mp F(-x, \xi, t), \quad \mbox{ and similarly for $H$ and $E$,} \nn\\
	&\wt{F}^{\pm}(x, \xi, t) = \wt{F}(x, \xi, t) \pm \wt{F}(-x, \xi, t), \quad \mbox{ and similarly for $\wt{H}$ and $\wt{E}$.}
\label{eq:charge-conjugation-relation}
\end{align}
The special pole $x_p$ is converted from \eq{eq:special-z1-pole} as
\beq
	x_p(\xi, z, \theta) = \xi \cdot \bb{ 2\rho(z) - 1 }
		= \xi \cdot \bb{ \frac{1 - z + \tan^2(\theta / 2) z}{1 - z - \tan^2(\theta / 2) z} }
\eeq
which crosses $[\xi, \infty) \cup (-\infty, -\xi]$ as $z$ goes from $0$ to $1$ and 
lies outside of the ERBL region.

%================================================================
\section{HARD COEFFICIENTS OF THE DY DILEPTON PRODUCTION PROCESS IN EQ.~\eqref{eq:dilepton}}
\label{app:dilepton}
%================================================================

As in Appendix \ref{app:diphoton}, the hard coefficient in \eq{eq:dilepton-factorize} for the DY dilepton process can be 
calculated in the SDHEP frame and expressed in terms of the helicity amplitudes of the leptons.
Parity and chiral symmetries constrain there to be only one independent amplitude,
\beq[eq:dilepton-helicity-structure]
	C^{\ell}_{\pm\mp} = \pm \frac{\N}{\hat{s}} C^{\ell}_0, \quad
	C^{\ell}_{\pm\pm} = 0 ,
\eeq
with $\N$ the same as in \eq{eq:diphoton-helicity-structure} and
\beq[eq:dilepton-coefs]
	2\xi C_0^{\ell} = i \sin\theta 
		\pp{ \frac{e_1}{1 - z_1 + i \epsilon} \frac{1}{z_2}
			- \frac{e_2}{z_1 + i \epsilon} \frac{1}{1 - z_2} }.
\eeq
Using the same notation as \eqs{eq:M-conv-short}{eq:GPD-DA-moments}, 
one determines the convolution of $C_0^{\ell}$ with the GPD and DA as
\begin{align}
	&\M_0^{[\wt{F}]} = -i \, D_0 \sin\theta
		\bb{ \frac{e_1 + e_2}{2} \wt{F}^-_0(\xi, t) + \frac{e_1 - e_2}{2} \wt{F}^+_0(\xi, t)	
			+ i \pi \pp{ \frac{e_1 + e_2}{2} \wt{F}^-(\xi, \xi, t) + \frac{e_1 - e_2}{2} \wt{F}^+(\xi, \xi, t) }
		}.
\label{eq:dilepton-M0}
\end{align}

%================================================================
\section{SHADOW GPD MODELS}
\label{app:sGPD}
%================================================================

The shadow GPDs used in this paper are the same as those in \refcite{Qiu:2023mrm}, 
except that we use
\begin{align}
	S_1(x, \xi) & = K_1 \, \xi^2 x (x^2-\xi ^2) (1-x^2)^{10} \pp{ 1 + a x^2 - \frac{29}{5} (2 a+27) x^4 + \frac{899}{35} (a+18) x^6}, \nn\\
	S_2(x, \xi) & = K_2 \, \xi^2 x (x^2-\xi ^2) (1-x^2)^{20} \pp{ 1 + a x^2 - \frac{49}{5} (2 a+47) x^4 + \frac{119}{5} (3a+94) x^6},
\end{align}
with $K_{1,2}$ the normalization factors, $a = -23$ for $S_1$ and $-38$ for $S_2$,
and we multiply the shadow $D$-term in \refcite{Qiu:2023mrm} by $2$.
Using the high powers for the factor $(1-x^2)$ signifies the shadow GPD magnitude near $|x| = \xi$,
as suggested by \fig{fig:kernel}.
	
\end{widetext}

%================================================================================
\bibliographystyle{apsrev}
\bibliography{reference}
%================================================================================

\end{document}